\documentclass[letter]{revtex4}
\usepackage{epsfig}
\usepackage{color}
\usepackage{xcolor}

\usepackage{tikz}
\usepackage{hyperref}

\definecolor{red}{rgb}{1,0,0}
\definecolor{blue}{rgb}{0,0,1}
\definecolor{black}{rgb}{0,0,0}

\newcommand{\be}{\begin{equation}}
\newcommand{\ee}{\end{equation}}
\newcommand{\ba}{\begin{eqnarray}}
\newcommand{\ea}{\end{eqnarray}}

\def\Ic{{\cal I}}
\def\Jc{{\cal J}}
\def\Lc{{\cal L}}
\def\Uc{{\cal U}}
\def\Oc{{\cal O}}
\def\Pc{{\cal P}}
\def\edot{\dot\epsilon}

\newcommand{\eq}[1]{\begin{align}#1\end{align}}

\newcommand\blfootnote[1]{%
  \begingroup
  \renewcommand\thefootnote{}\footnote{#1}%
  \addtocounter{footnote}{-1}%
  \endgroup
}

\usepackage{amsmath}
\newlength{\arrow}
\begin{document}

\title{Unified Theory of  Inertial Granular Flows  and Non-Brownian Suspensions}

\author{ E. DeGiuli${}^1{}^\dagger$\blfootnote{${}^\dagger$ These authors contributed equally to this work.}, G. D\"uring${}^2{}^\dagger$, E. Lerner${}^{1,3}{}^\dagger$,  and M. Wyart${}^1$}
\affiliation{$^1$New York University, Center for Soft Matter Research, 4 Washington Place, New York, NY, 10003, \\
${}^2$ Facultad de F\'isica, Pontificia Universidad Cat\'olica de Chile, Casilla 306, Santiago 22, Chile
\\
${}^3$ Institute for Theoretical Physics, Institute of Physics, University of Amsterdam,
Science Park 904, 1098 XH Amsterdam, The Netherlands 
}


\begin{abstract}
Rheological properties of dense flows of hard particles are singular as one approaches the jamming threshold where flow ceases, both for aerial granular flows dominated by inertia, and for over-damped suspensions. Concomitantly, the lengthscale characterizing velocity correlations appears to diverge at jamming.  Here we introduce a theoretical framework that proposes a tentative, but potentially complete scaling description of stationary flows.  Our analysis, which focuses on frictionless particles, applies {\it both} to suspensions and inertial flows of hard particles. We compare our predictions with the empirical literature, as well as with novel numerical data. Overall we find a very good agreement between theory and observations, except for frictional inertial flows whose scaling properties clearly differ from frictionless systems. For over-damped flows, more observations are needed to decide if friction is a relevant perturbation or not. Our analysis makes several new predictions on microscopic dynamical quantities that should be accessible experimentally. 

\end{abstract}


\maketitle

\section{Introduction}
Microscopic description  of particulate materials such as grains, emulsions or suspensions is
complicated by the presence of disorder, and by the fact that these systems are often out-of-equilibrium.
One of the most vexing problems is how these materials transition between a flowing and a solid phase.  
When this transition is driven by temperature, it corresponds  to the glass transition where a liquid becomes
 a glass, an amorphous structure that cannot flow on experimental time scales. Here we focus instead on athermal systems driven by an imposed stress, such as granular flows, and consider both the case where inertia is important (such as in aerial granular flows) or not (such as over-damped suspensions). We focus primarily on the case of hard particles. 

 Empirical constitutive relations have been proposed to describe such dense flows in the limit of hard particles \cite{MiDi04,Cruz05,Jop06}. Two important dimensionless quantities are the packing fraction $\phi$ and the stress anisotropy $\mu\equiv \sigma/p$  (also called the effective friction), where $\sigma$ is the applied shear stress and $p$ the pressure carried by the particles. For inertial flow, dimensional analysis implies that both quantities can only depend on the strain rate $\dot\epsilon$, $p$, the particle diameter $D$ and the mass density of the hard particles $\rho$ via the inertial number ${\cal I}\equiv \dot \epsilon D \sqrt{\rho/p}$. One finds empirically that the constitutive relations $\mu({\cal I})$ and $\phi({\cal I})$ converge to a constant as ${\cal I}\rightarrow 0$, corresponding to the jamming transition where flow stops. We define $\mu(0)\equiv \mu_c$ and $\phi(0)\equiv\phi_c$, which are system-specific and will depend on particle shape, poly-dispersity, friction coefficient, etc. 
 Near jamming, the constitutive relations are observed to be singular with:
\ba
\label{0}
\delta \mu\equiv \mu({\cal I})-\mu_c\propto {\cal I}^{\alpha_\mu} \\
\delta \phi\equiv \phi_c-\phi({\cal I})\propto {\cal I}^{\alpha_\phi} \label{01}
\ea
As jamming is approached the dynamics becomes increasingly  correlated in space \cite{Pouliquen04,Olsson07}. By considering the dominant decay \cite{During14} of the velocity correlation function, one can define a length scale $\ell_c$: 
\be
\label{02}
\ell_c \sim {\cal I}^{-\alpha_\ell}
\ee
Similar dimensional arguments have been made for dense suspensions of non-Brownian particles  \cite{Lemaitre09b,Boyer11}. 
In that case the  relevant dimensionless number is the viscous number ${\cal J}= \eta_0\dot\epsilon/p$, where $\eta_0$ is the viscosity of the solvent. Empirically one finds similar relations:
\ba
 \label{03}
\delta \mu\equiv \mu({\cal J})-\mu_c\propto {\cal J}^{\gamma_\mu} \\
\delta \phi\equiv \phi_c-\phi({\cal J})\propto {\cal J}^{\gamma_\phi} \label{04}\\
\ell_c \sim {\cal J}^{-\gamma_\ell}  \label{05}
\ea
These relations imply that the viscosity  $\eta=\sigma/\edot$ of the suspension diverges as jamming is approached. Indeed Eq.(\ref{03}) implies that $\sigma\sim p$ near jamming (in our scaling arguments below we may thus exchange freely $\sigma$ and $p$), so that ${\cal J}\propto \eta_0/\eta$. Eq.(\ref{04}) then implies that:
\be
\label{06}
\frac{\eta}{\eta_0}\propto (\phi_c-\phi)^{-1/\gamma_\phi}
\ee
 When both viscosity and inertia are present, a transition from viscous to inertial flow occurs as strain rate $\edot$ is increased at fixed volume fraction \cite{Fall10,Trulsson12,Vagberg14}. This defines a cross-over strain rate
\eq{
\label{epsco}
\edot_{v\rightarrow i} \propto \frac{\eta_0}{\rho D^{2}} (\phi_c - \phi)^{\gamma_{\edot}},
}
where the prefactors follow from a dimensional analysis.

Empirical values found for the exponents in Eqs.(\ref{0}-\ref{06}) are reported in Table 1. They seem not to depend on dimension, which we thus did not report in our table. In the case of inertial flow they appear to depend on the presence of friction, whereas for suspended particles exponents appear to be similar with and without friction. In this work we focus on frictionless particles, and discuss open questions on frictional systems in the conclusion. 

\begin{figure*}[b!] 
\includegraphics[width=0.50\textwidth,clip]{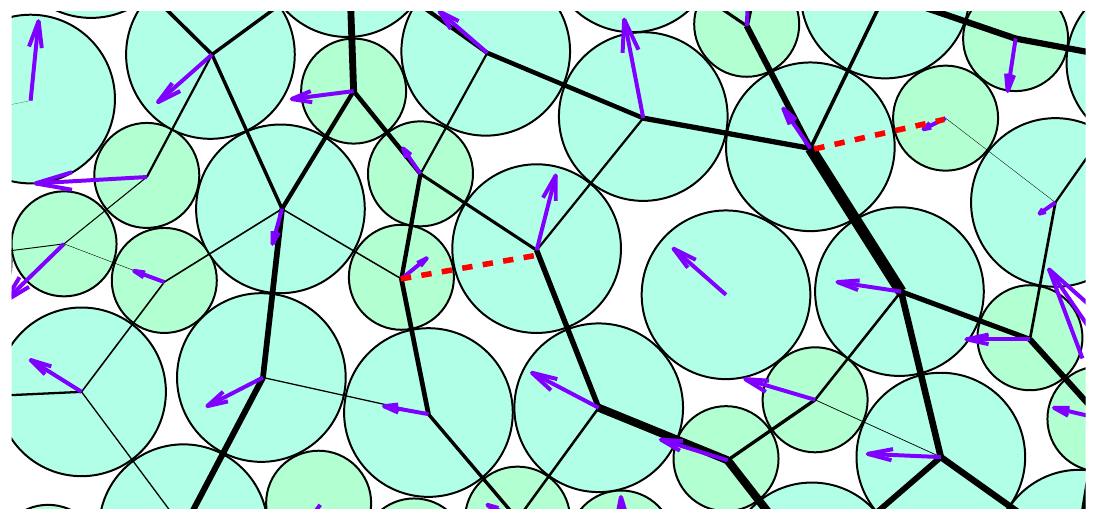}
\caption{(Color online) Illustration of solid destabilization: several weak contacts, indicated by red dashed lines, are opened. This induces a space of extended, disordered floppy modes, one of which is shown (arrows). Line thickness indicates force magnitude in the original, stable solid.}\label{packing}
\end{figure*}

\begin{figure*}[t!] 
\includegraphics[viewport = 30 75 755 430, width=0.6\textwidth,clip]{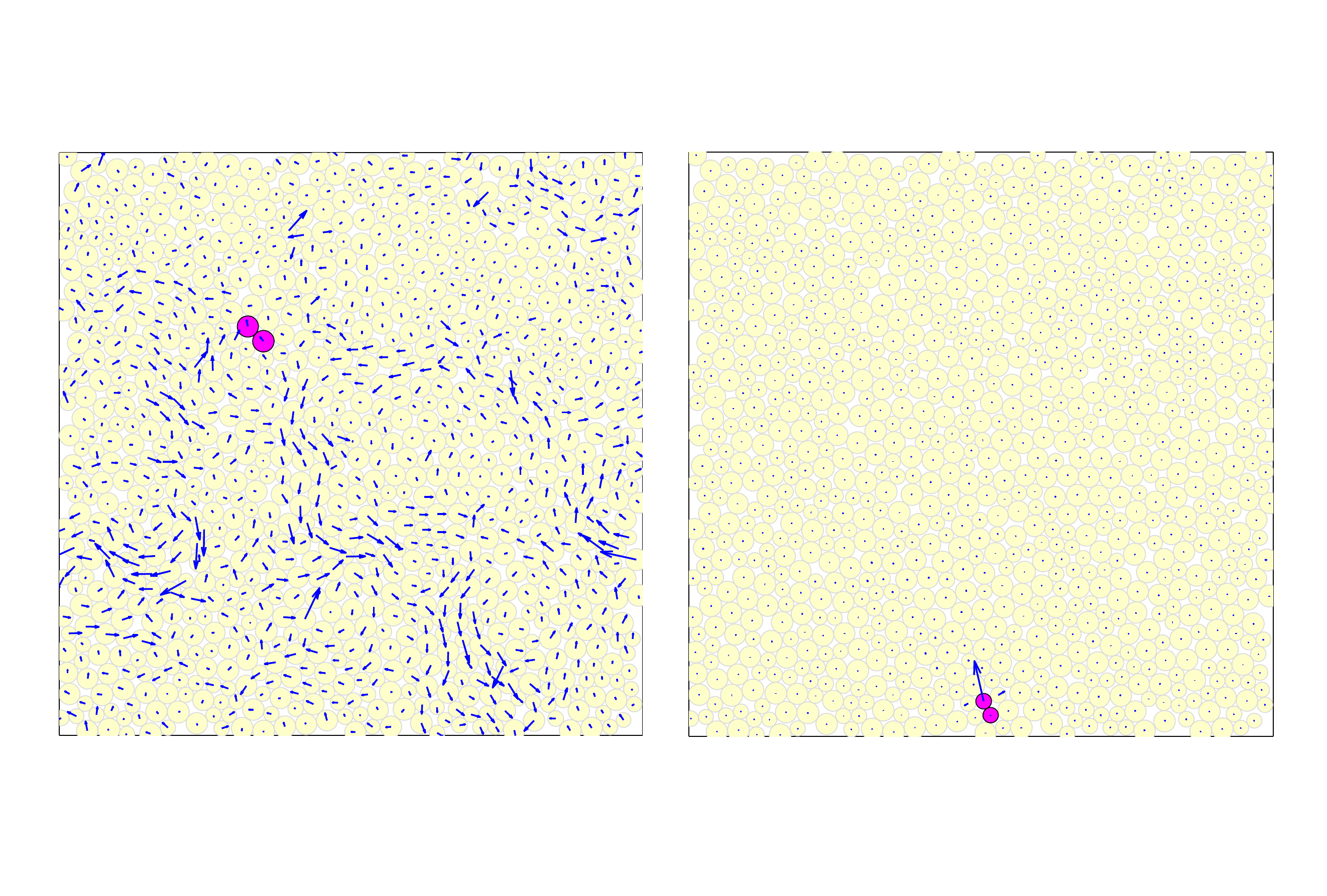} 
\caption{(Color online) Extended {\it vs} localized contacts. When a contact is opened from an isostatic packing, the resulting deformation (arrows) can either be {\it extended}, as shown at left, or {\it localized}, as shown at right. Localized contacts are more numerous, but only extended contacts couple strongly to an imposed shear stress. Reproduced from \cite{Lerner13a} by permission of The Royal Society of Chemistry (RSC). }\label{twomodes}
\end{figure*}

Currently there is no accepted microscopic theory describing quantitatively these singular behaviors, in particular Eqs.(\ref{0}-\ref{06}). Various works \cite{Kumaran06,Berzi08,Seguin11} propose to describe dense flows by a perturbation around the dilute limit $\phi\rightarrow 0$. In the case of dense suspensions, this corresponds to extending the work of Einstein and Batchelor, who computed the first corrections to the viscosity at larger density. For  dry  granular flows, this corresponds to an extension of kinetic theory ({\it a priori} valid in the gas phase) to the dense regime. However, observations support that as jamming is approached, particles form an extended network of contacts, and that the stress is dominated by contact forces \cite{Cruz05, Peyneau08,Boyer11}. In this work we propose a framework to describe flow in such situations. 

We attack the problem  in two steps. First, we isolate the  microscopic quantities that control flow. Then, we compute the scaling properties of these quantities  by performing a perturbation around the solid phase. The idea is to consider the solid in the critical state, i.e. carrying the maximal anisotropy possible $\mu=\mu_c$, corresponding to a packing fraction $\phi=\phi_c$. Next, one adds an additional kick to the system, corresponding to a small  additional stress anisotropy $\delta \mu$. As a result, some contacts between particles will open, forces will be unbalanced, and the system will start to flow (see Fig. \ref{packing}).  Our key assumption is  that flowing configurations are similar to a solid that is thus destabilized. As we will see, this approach  enables us to propose a full scaling description of the problem, and to predict the exponents entering Eqs.(\ref{0}-\ref{06}) in good agreement with observations  in the absence of static friction. Moreover, our approach predicts several other properties singular near jamming: the speed of the particles, the strain scale beyond which a particle loses memory of its velocity, and the coordination of the contact network. The first two quantities are accessible experimentally, and provide an additional experimental test of our views. 

In \cite{Lerner12} three of us have already proposed to perform a perturbation around the solid. However, this argument was limited to over-damped suspensions, and did not predict the scaling relations entering in the constitutive relations Eqs.(\ref{0}-\ref{06}). Moreover, a key aspect of the argument turned out to be incorrect: it was assumed that when an additional stress anisotropy is imposed, the contacts carrying the smallest forces open. This assumption led to a scaling description for the viscosity and several microscopic quantities in terms of an exponent $\theta_\ell$, characterizing the distribution of weak forces in packings. Later it was realized that only a vanishingly small fraction of weak contacts are significantly coupled to external stresses \cite{Lerner13a}. We call them {\it extended} contacts, because perturbing such contacts mechanically lead to a spatially extended response in the system, as shown in Figure \ref{twomodes} \cite{Lerner13a,DeGiuli14b}. In a packing only those contacts lead to plasticity when stress is increased, or when a shock (say a collision) occurs in the bulk of the material  \cite{DuringNS,Lerner13a}. The density of extended contacts as a function of the rescaled force $\tilde{f}=f/p$ in the contact follows:
 \be
 \label{08}
 P({\tilde f})\propto {\tilde f}^{\theta_e}.
 \ee
Numerically it is found that $\theta_e\approx 0.44$ both in two and three dimensions \cite{Lerner13a,DeGiuli14b}, suggesting that this quantity may be independent of dimension. Moreover, its value does not depend on the preparation protocol of the isostatic state: up to error bars, equal values are found from compression of hard spheres \cite{Lerner13a}, shear-jammed hard disks \cite{Lerner12}, and decompression of soft spheres \cite{DeGiuli14b,Charbonneau15} \footnote{We include here works where $\theta_e$ was not directly measured, but inferred from the force distribution exponent $\theta_\ell$ by the marginal stability relation $\theta_e=2\theta_\ell/(1-\theta_\ell)$. See \cite{DeGiuli14b}.}. The exponent $\theta_e$ can be shown to control the stability of the solid phase \cite{Wyart12,Lerner13a,Muller14}. Recently replica calculations in infinite dimension on the force distribution \cite{Charbonneau14,Charbonneau14a,Charbonneau15} led to the prediction \cite{DeGiuli14b}:
\be
\theta_e=0.423...,
\ee  
within the error bar of our measurements. In our proposed scaling description all exponents can be expressed in terms of $\theta_e$, in particular:
\ba
\alpha_\mu&=&\alpha_\phi=\gamma_\mu=\gamma_\phi= \frac{3+\theta_e}{8+4\theta_e}\approx 0.35 \\
\gamma_{\edot} &=& \frac{8+4\theta_e}{3+\theta_e} \approx 2.83 \\
\alpha_\ell&=& \gamma_\ell=   \frac{1+\theta_e}{8+4\theta_e}\approx 0.15  
\ea
Empirically it was noticed that $\gamma_\mu=\gamma_\phi$ and that $\alpha_\mu=\alpha_\phi$,
which our arguments rationalize. 

\subsection{General approach}
We argue that several dimensionless quantities that characterize the microscopic dynamics under flow critically affect rheological properties. 
As jamming is approached, the assembly of particles acts as a lever: due to steric hindrance, the typical relative velocity between adjacent particles $V_r$ becomes much larger than the characteristic velocity $\dot \epsilon D$ where  $\dot \epsilon$ is the strain rate and $D$ the mean radius of the particles \cite{Lerner12a,Andreotti12}. We thus define the amplitude of this lever effect ${\cal L}$ as:
\be
\label{1}
{\cal L}=\frac{V_r}{\dot \epsilon D}.
\ee
Another fundamental quantity, particularly relevant for inertial flow, is the strain scale $\epsilon_v$ beyond which a particle loses memory of its direction relative to its neighbors. $\epsilon_v$ can be extracted from the decay of the  autocorrelation function $\langle V^\alpha_r(0) V^\alpha_r(\epsilon)\rangle$, where the average is made over all pairs of adjacent particles $\alpha$. A similar quantity was extracted numerically in  \cite{Olsson10a}.
As the packing fraction $\phi$ increases toward jamming, collisions are more frequent per unit strain (due to the increase of relative particle motion ${\cal L}$), and each collision affects the motion of the particles on a growing length scale. These two effects implies that $\epsilon_v$ vanishes rapidly near jamming.

We now argue that dissipation is entirely governed by $\Lc$ in overdamped suspensions, and by both $\Lc$ and $\epsilon_v$ in inertial flows.
In both cases the power injected into the system at the boundaries, which is simply ${\cal P}=\Omega \sigma \dot\epsilon$ at constant volume, must be dissipated in the bulk.  

In a dense suspension we expect dissipation to be governed by local mechanisms such as lubrication. Lubrication forces are singular for the ideal case of perfectly smooth spheres, but not for rough particles where they must be cut off. Thus the viscous force exchanged by two neighboring particles  must dimensionally follow $F\sim\eta_0 V_r D^{d-2}$, leading to a power dissipated ${\cal P}/N=C \eta_0 V_r^2 D^{d-2}$  where $d$ is the spatial dimension, $C$ is a dimensionless constant that depends on the particle shape and roughness, and $N$ is the number of particles. Equating the power dissipated to the power injected, one gets that for a given choice of particles:
\be
\label{2}
\frac{\eta}{ \eta_0}\propto 1/{\cal J}\propto {\cal L}^2
\ee
implying that the divergence of viscosity  is governed by  ${\cal L}$.  This  result holds by construction in simple models of dissipation in suspension flows \cite{Lerner12a,Andreotti12,Vagberg14b}.

Concerning inertial flows, we  suppose that the restitution coefficient characterizing a collision between two particles is smaller than one, and that collisions dominate dissipation. Then each time two neighboring particles change relative direction, a finite fraction of their relative kinetic energy $E_c\sim M V_r^2$  must be dissipated, where $M$ is the particle mass. Then the total power dissipated must follow ${\cal P}\propto N \dot \epsilon E_c /\epsilon_v$. Using Eq.(\ref{1}) and balancing power injected and dissipated, one gets $\sigma/(\dot\epsilon^2 D^2 \rho )\sim \Lc^2/\epsilon_v$ where $\rho$ is the mass density of the particles, so that the inertial number ${\cal I}$ follows:
\be
\label{4}
{\cal I}\sim \frac{\sqrt{\epsilon_v}}{{\cal L}}
\ee
To our knowledge Eq.(\ref{4}) has not been proposed before, and could be tested empirically.

\begin{figure*}[t!] 
%
%
\includegraphics[width=0.40\textwidth,clip]{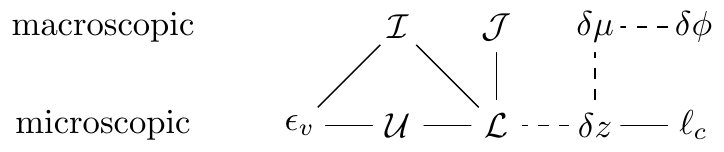}
\caption{Outline of logical relationships between main macroscopic and microscopic quantities, showing the key role that $\Lc$ and $\delta z$ have in relating control parameters to the shear rate. Dashed lines indicate arguments that use an ansatz of flowing configurations being similar to destabilized isostatic ones, whereas solid lines indicate arguments independent of this assumption. }\label{diagram}
\end{figure*} 

\subsection{Organization of the manuscript}
To obtain a complete description of flow, one must therefore express $\Lc$ and $\epsilon_v$ in terms of control parameters such as $\delta \mu$ or $\delta \phi$.  To achieve this goal, we make the assumption that the contact network of configurations in flow is similar to that of jammed configurations at $\mu_c$ immediately after increasing the stress anisotropy by $\delta \mu>0$. The coordination $z$ of the network of contacts is a key microscopic quantity that distinguishes flowing from jammed configurations.  At jamming the coordination is just sufficient to forbid motion, corresponding to $z_c=2d$ for frictionless spheres \cite{Tkachenko99,Moukarzel98,Roux00}.  As illustrated in Fig.(\ref{packing}), the kick of amplitude $\delta \mu$ opens a fraction $\delta z\equiv z_c-z$ of the contacts, allowing collective motions of the particles for which particles do not overlap, but simply stay in contact, the so-called called {\it floppy modes}. In Section \ref{S1}, we argue based on simple geometrical considerations that the lever amplitude is directly related to the density of floppy modes $\delta z$, and obtain:

\be
\label{001}
\Lc\sim \delta z^{-(2+\theta_e)/(1+\theta_e)}
\ee
In Section \ref{S2} we consider the evolution of contact forces with stress anisotropy in a jammed packing, and argue that the number of contacts that open follows:
\be
\label{002}
\delta z \sim \delta \mu^{(2+2\theta_e)/(3+\theta_e)}
\ee
Jointly Eqs.(\ref{001},\ref{002}) predict a relationship between level amplitude and stress anisotropy: 
\be
\label{6}
{\cal L}\sim  \delta \mu^{-(4+2\theta_e)/(3+\theta_e)}\sim\delta\mu^{-1.41}
\ee 
Eqs.(\ref{2},\ref{6}) lead to a prediction for the exponent $\gamma_\mu$ entering in the constitutive relation $\mu({\cal J})$. 
Together with previous results showing that $\ell_c\sim 1/\sqrt{\delta z}$ \cite{During13,During14}, one obtains expressions for $\gamma_\ell$ and $\alpha_\ell$,
corresponding to:
\be
\label{6bis}
\ell_c\sim \delta \mu^{-(1+\theta_e)/(3+\theta_e)}\sim \delta\mu^{-0.41}
\ee
both for inertial and viscous flows. 
In Section \ref{S3}  we investigate the characteristic strain scale $\epsilon_v$ at which velocities decorrelate. We compute the decay of stress occurring in between collisions at fixed packing fraction, as well as the positive jump of stress that occur when new contacts are formed. Stationarity then implies that these two quantities must be equal in average, leading to the prediction that in steady state:
\be
\label{7}
\epsilon_v \sim 1/\Lc^2.
\ee
Together with Eqs.(\ref{4},\ref{6}) this result leads to a prediction for the exponent $\alpha_\mu$ characterizing the constitutive relation $\mu({\cal I})$.
One missing link to obtain a full scaling description of the problem is how the packing fraction depends on other control parameters. In section \ref{S4} we make the additional assumption that isotropic packings of frictionless particles in the thermodynamic limit have a finite (although presumably small) dilatancy.  We show that this hypothesis implies the scaling relation $\delta\phi\sim \delta \mu$, known to agree well with observations. This result enables us to predict the exponents $\alpha_\phi$ and $\gamma_\phi$ entering the constitutive relation for $\phi({\cal I})$ and $\phi({\cal J})$, leading to a complete scaling description of rheological properties near jamming for frictionless particles. In particular the divergence of viscosity with packing fraction in suspensions is expected to follow:
\be
\label{8bis}
\frac{\eta}{\eta_0}\sim (\phi_c-\phi)^{-(8+4\theta_e)/(3+\theta_e)}\sim (\phi_c-\phi)^{-2.83}
\ee
An outline of the logical relationships between the main macroscopic and microscopic quantities is shown in Fig. \ref{diagram}. 

In Section \ref{S4a} we study the transition from viscous to inertial flow. In Section \ref{S6} we compare our results with previous empirical and numerical observations. Overall we find a very good agreement between observations and predictions for frictionless particles. We conclude by discussing open problems, such as the influence of friction, which appears to change exponents for inertial flows.

\section{Lever effect ${\cal L}$ and coordination}
\label{S1}

 To compute how the lever effect ${\cal L}$  depends on the deficit in coordination $\delta z$, we again 
 consider an anisotropic jammed packing with $z=z_c$,  and remove the $\delta z$ weakest  extended contacts (our procedure is equivalent to instantaneously eroding the surfaces of the particles making those contacts, allowing particles to flow toward each other). The system can now flow along {\it floppy modes}, i.e. collective motions along which particles in contact remain at the same distance. These floppy modes pervade the system \cite{Lerner12a}.

Before the contacts were removed, forces were balanced on every particle. 
 A formal way to write force balance is the virtual work theorem, recalled in Appendix A. 
It states that {\it for any displacement field} $\{{\delta \vec R}_i\}$, the work of external forces is equal to the work of contact forces:
\be
\label{14}
\sum_i {\vec F^{ext}_i}\cdot  \delta {\vec R}_i= - \sum_{ ij} f_{ij} \delta r_{ij}
\ee
where $f_{ij}>0$ is the contact force in the contact $ij$, and   $\delta r_{ij}$  is the change of distance between particles in contact, $\delta r_{ij}\equiv (\delta{\vec R}_j-\delta {\vec R}_i)\cdot {\vec n}_{ij}$ where ${\vec n}_{ij}\equiv ({\vec R}_j-{\vec R}_i)/||{\vec R}_i-{\vec R}_j||$ and $\vec{R}_i$ is the position of particle $i$. We can use Eq.(\ref{14}) for the floppy modes that would appear if contacts were removed. For floppy modes, $\delta r_{ij}=0$ except for the fraction $\delta z$ of the contacts removed for which $\delta r_{ij}<0$. 
On the other hand external forces are only present at the boundaries, and the left-hand side of Eq.(\ref{14})  corresponds to the work of the applied stress, which for a simple shear reads $\Omega \sigma  \delta \epsilon$, where $\Omega$ is the volume, and $\delta\epsilon$ is shear strain. Overall we get:
\be
\label{15}
\Omega \sigma \delta \epsilon= -\sum_\alpha f_{\alpha} \delta r_{\alpha}\sim N \delta z \delta r  f(\delta z)
\ee
where the sum is on the $N\delta z$ contacts that were removed, labeled by $\alpha$. In Eq.(\ref{15}) we estimated this sum by introducing the characteristic magnitude of displacements in a floppy mode, $\delta r$, and the characteristic force, $f(\delta z)$, of the contacts removed. It satisfies:
 \be
 \label{16}
 \int_0^{f(\delta z)/p}P(f'/p)d (f'/p)\sim\delta z 
 \ee
leading to $f(\delta z)\sim p \delta z^{1/(1+\theta_e)}$. Together with Eq.(\ref{15}) and using that $p\sim \sigma$ near jamming, we get:
\be
\label{17}
{\cal L}\sim \frac{\delta r}{\delta \epsilon}\sim \delta z^{-(2+\theta_e)/(1+\theta_e)}
\ee
Eq.(\ref{17}) was first derived by some of us for some specific models \footnote[0]{ In the Affine Solvent Model \cite{Lerner12,Lerner12a}, the divergence of ${\cal L}$ is associated to the presence of a vanishing eigenvalue in the operator ${\cal N}$ that controls flow \cite{Lerner12a}. The variational argument \cite{Lerner12} led to  ${\cal L}\sim\delta z^{-(3+\theta)/(2+2\theta)}$, a result that does not hold when extended contacts are included. The variational argument of \cite{Lerner12}  can be improved in that case by lowering the force unbalance via applying dipoles where contacts are broken, which can be shown to lead to Eq.(\ref{17}).}, a result that will be published elsewhere \cite{DuringNS}.   It is important to note that our derivation should hold true for any deficit in coordination $\delta z$, up to the smallest values it can take, i.e. $\delta z\sim 1/N$ (corresponding to ${\cal O}(1)$ contacts  removed). The same situation occurs for soft particles above jamming, where it was argued that the shear modulus $G$ vanishes as $G\sim \delta z$ with a scaling that holds up to $\delta z\sim1/N$ \cite{Wyart05b}, as confirmed numerically \cite{Goodrich12}.  In both cases this  behavior traces back to the fact that near {\it isostaticity} (i.e. $z=z_c$), the physics is governed by counting arguments (for example the summation in Eq.(\ref{15})), which apply irrespective of the magnitude of $\delta z$. Henceforth we assume that it is the case for quantities of interest. 

\section{Relation between coordination and stress anisotropy}
\label{S2}
We now seek to determine the relationship between the coordination deficit $\delta z$ and the increment of stress anisotropy $\delta \mu$. In particular we define the exponent:
\be
\label{10}
\delta \mu\sim \delta z^{y_\mu}
\ee
To compute the exponent $y_{\mu}$ we consider jammed configurations, and estimate the increment of stress anisotropy  $\delta \mu_N$ required to open one contact, thus corresponding to $\delta z\sim 1/N$.  For simplicity we consider isotropic packings, and use the fact that at jamming the shear modulus of soft particles is of order $G\sim 1/N$ \cite{Wyart05b,Wyart10b,Goodrich12}. We show in Appendix B that this argument is unchanged for anisotropic packings. In Appendix C, we provide another argument which does not assume the behavior of $G$, and considers strictly hard particles. It uses simple geometrical considerations and an assumption on the randomness of contact forces in packings.  

Consider a packing of hard particles. For our purpose it is convenient to approximate such hard particles by soft harmonic particles of  stiffness $k$, in the limit where they are not deformed, i.e.  $p/k \rightarrow 0$. A shear modulus can then be defined, that follows $G\sim k/N$  \cite{Wyart05b,Wyart10b,Goodrich12}. 
If a shear stress increment $\delta \sigma=p \;\delta\mu$ is imposed at the boundaries, contact forces will change by some characteristic amount $\delta f$, leading  to a characteristic change of energy $\delta f^2/k$ in  contacts.  The energy per particle $\delta E$ stored in the system is thus of the order $\delta f^2/k$. By definition of the shear modulus, one must also have $\delta E\sim \delta \sigma^2/G \sim N \delta \sigma^2/k\sim Np^2\delta \mu^2/k$. Comparing these expressions we obtain
\be
\label{8}
\delta f\sim p \delta \mu\sqrt{N}. 
\ee
The first  contact opens when $\delta f$ becomes of order of the smallest force in the system, $f_s$, which satisfies
 \be
 \label{12}
 \int_0^{f_s/p}P(f'/p)d (f'/p)\sim 1/N,
 \ee
 where $P(x)\sim x^{\theta_e}$, leading to $ f_s\sim p N^{-1/(1+\theta_e)}$. Equating this expression with $\delta f$ implies
 \be
 \label{4bis}
 \delta\mu_N\sim N^{-1/2-1/(1+\theta_e)},
 \ee
a result in excellent agreement with the numerics of \cite{Combe00}. Comparing Eqs.(\ref{10},\ref{4bis}) for $\delta z=1/N$ we get
 \be
 \label{13}
 y_\mu=\frac{3+\theta_e}{2(1+\theta_e)}.
 \ee

\section{Strain scale in flow}
\label{S3}

\subsection{Connection between lever amplitude and force unbalance}

It is useful to realize that the lever amplitude ${\cal L}$ directly connects to the ability of the contact network to  balance forces. Denoting $f$ the characteristic amplitude of the force in the contacts between particles, and $F$ the characteristic amplitude of the sum of these forces on one particle, we define the dimensionless quantity:
\be
\label{5}
{\cal U}=\frac{F}{f}.
\ee
If forces are balanced as in a static granular assembly, then obviously ${\cal U}=0$, a limit that is reached continuously as jamming is approached. Using the fact that a flowing configuration has deformation modes permeating the system, it is possible to show (see Appendix A) that independent of the presence of inertia, the power done in deformation can be written as
\eq{ \label{W}
{\cal P} = \sum_i \vec{F}_i \cdot \vec{V}_i, 
}
where $\vec{F}_i$ is the vectorial sum of the contact forces on particle $i$, and $\vec{V}_i$ its  velocity. 
It is useful to decompose $\vec{V}_i$ into the motion of the particle relative to its neighbors $\vec{V}^r_i$,  and the convection $\vec{V}^c_i$ of the region of the system surrounding particle $i$-- which includes in particular the mean velocity of the flow, the so-called affine velocity. Galilean invariance and isotropy of space imply that the local force and the affine velocity are not correlated on average, and we expect in general that 
 $\langle \vec{F}_i \cdot \vec{V}^c_i\rangle\approx0$ \footnote[1]{This relation is certainly an equality for the affine velocity. However, it is possible that long wavelength displacement modes contribute significantly to non-affine displacements in models where  dissipation is purely due to the relative motion between particles. Such long wavelength displacements contribute to the local convection of  particles. However we expect that these are weakly correlated to
the local force, which is determined by the local packing structure.}.
However for generic dissipation mechanisms, $\vec{F}_i$ and $\vec{V}^r_i$ are correlated, so that on average $\vec{F}_i \cdot \vec{V}^{r}_i > 0$. 
Eq.\eqref{W} thus implies $\Pc \sim N F V^{r} \sim N \Uc f \Lc \edot D$. Recall that the power injected at the boundary is $\Pc =\Omega \sigma \dot\epsilon$. Using  $f \propto \sigma D^{d-1}$ we get $\Pc \sim \Omega f \dot\epsilon D^{1-d}$. Comparing these two expressions for the dissipated power leads to
\eq{
\label{5bis}
\Lc\sim 1/\Uc.
}

\subsection{Decay of {\cal L} at fixed coordination}
 
The force unbalance ${\cal U}$ increases with strain as the particles are convected with no contact creation at fixed volume, as long as contact forces are positive (which is always true for purely repulsive particles). This effect can be illustrated in the simple example of a nearly straight line of connected rigid rods. ${\cal U}=0$ if the line is completely straight, but increases as the line is compressed and forms a zig-zag. However if the line is pulled, contact forces are then negative, and ${\cal U}$ decreases toward zero. The explanation for this fact stems from a simple geometrical consideration: as one particle $i$ moves, the direction of its contacts ${\vec n}_{ij}$  tends to turn away from the direction of motion. 
Since the resultant of the contact force  is ${\vec F}_i=-\sum_{ij} f_{ij} {\vec n}_{ij}$, the projection of the force on the direction of motion increases if contact forces are positive. Because for generic dissipation mechanism, one expects that ${\vec F}_i$ and the particle velocity with respect to its neighbors $\vec{V}^{r}_i$  are correlated,  the norm of the unbalanced force grows in average. This effect is proportional to the change of orientation of the contact, itself proportional to $\vec{V}^{r}_i$ and thus to the lever amplitude ${\cal L}$, leading to:
%
%
\be
\label{19}  
\frac{\partial {\cal U}}{\partial \epsilon}\sim {\cal L},
\ee
which simply means that the faster the motion, the faster forces become unbalanced. Eqs.(\ref{19},\ref{5bis}) then imply
\be
\label{20}
\frac{\partial {\cal L}}{\partial \epsilon}\sim -{\cal L}^3.
\ee
\begin{figure*}[b!] 
\includegraphics[viewport=70 240 505 425,width=0.50\textwidth,clip]{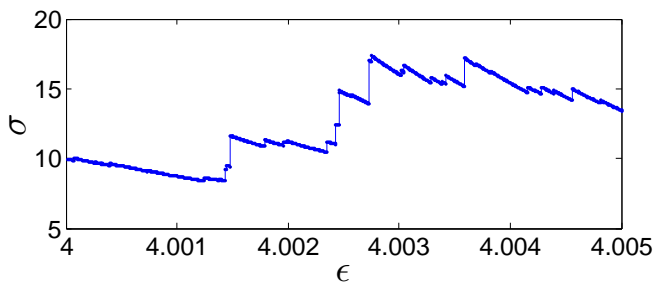}
\caption{(Color online) Typical stress {\it vs} strain curve for flow of rigid particles, showing intervals where stress relaxes smoothly, punctuated by instantaneous collisions (vertical segments). }\label{collision}
\end{figure*}

 \subsection{Consequence of stationarity}
 
Eq.(\ref{20}) was argued for in \cite{During14} for a specific model of suspension flow, where it was noticed that Eq.(\ref{20}) indicates the presence of a characteristic strain scale $\epsilon_\eta\sim 1/{\cal L}^2$, for which the viscosity would decay by a finite amount if no contacts were created. At fixed volume, this effect leads to a decrease of stress between collisions apparent in simulation, see Fig.(\ref{collision}). We now use a finite-size scaling argument to argue that the velocities decorrelate on the same strain scale, i.e $\epsilon_v\sim \epsilon_\eta$ (we will give elsewhere an alternative derivation of this result for a specific model of flow \cite{DuringNS}).


To do so we consider a system with  only a few floppy modes, corresponding to $\delta z\sim1/N$. As discussed above we still expect scaling relations to hold in that situation. 
In such a system, the lever amplitude ${\cal L}$ decreases in between collisions, as described by Eq.(\ref{20}).   However, a collision can decrease the number of floppy modes by adding one contact, and thus increases ${\cal L}$. Since the relative fluctuations of the number of floppy modes in a subsystem are of order one, and using that this number and  ${\cal L}$ have a power-law relation (\ref{17}), a collision must locally change ${\cal L}$ by $\Delta {\cal L}\sim {\cal L}$.  In a stationary state, any such increase of ${\cal L}$ must be compensated by the decrease of ${\cal L}$ in between collisions. According to Eq.(\ref{20}), this can occur only if the collision rate in the subsystem is $\epsilon_\eta\sim 1/{\cal L}^2$. Collisions result in a change in the nature of floppy modes, and must therefore decorrelate the  particles velocities by some finite fraction \footnote[2]{To show this it is sufficient to prove that in a system of size $N$ with one floppy mode, replacing a contact $\alpha$ by some new contact $\beta$ decorrelates the floppy mode by some finite amount, i.e. $\langle \delta R_\alpha|\delta R_\beta\rangle$ does not converge to one $1$ as $N\rightarrow\infty$, where  $| \delta R_\alpha\rangle$ ($| \delta R_\beta\rangle$) is the normalized floppy mode before (after) the contact was changed. Consider the system with the two contacts $\alpha$ and $\beta$ removed. Floppy modes then form a vector space of dimension 2. Adding back the contact $\alpha$ corresponds to choosing a random direction in this space of two dimensions, defining $| \delta R_\alpha\rangle$.  Instead if  the contact $\beta$ is added another direction $| \delta R_\beta\rangle$ is defined. Two random directions is a space of dimension two are generically not collinear, but instead make some angle which is ${\cal O}(1)$.}, implying  $\epsilon_v\sim \epsilon_\eta$ and therefore Eq.(\ref{7}). This prediction agrees very well with numerical models of suspension flows \cite{Olsson10a}, as we shall confirm with new data in Section \ref{S6}.

\section{Packing fraction}
\label{S4}
We now provide a finite size scaling argument  supporting that $\delta \phi\sim \delta \mu$. 
In Eq.(\ref{4bis}) we computed the increment of stress anisotropy required to open one contact in a jammed solid,
$\delta \mu_N$. We will argue that on average, opening one contact triggers an avalanche leading to a mean 
change of packing fraction:
\be
\label{21}
\langle \delta \phi_N\rangle \sim \delta \mu_N.
\ee
 From this result we argue that $\delta \phi\sim \delta \mu$ as follows. Consider an infinitely large packing at $\mu_c$, and increase the stress anisotropy by some $\delta \mu>0$. As argued in Eqs.(\ref{10},\ref{13}) this will open $\delta z$ contacts. We may next cut the system in subsystems of volume $\Omega_{FS}\sim 1/\delta z$. Following the opening of $\sim \Oc(1)$ contact, each subsystem will change its packing fraction by some amount $\delta \phi_{\Omega_{FS}}$. The average of this quantity will determine the change of packing fraction $\delta \phi$ in the entire system. According to Eq.(\ref{21}) this average is simply $\delta \mu$, leading to the desired result $\delta \phi\sim \delta \mu$.

To prove Eq.(\ref{21}) we consider the plasticity of a stable packing of $N$ hard particles, i.e. with $\mu<\mu_c$. Let us assume that the stress anisotropy cycles adiabatically between $-\mu_1$ and $\mu_1$, where $\mu_1$ is smaller than, but of order, $\mu_c$. We expect that in the thermodynamic limit, the packing fraction will be minimal at $\mu=0$, and we shall assume that it rises to a finite (although presumably numerically small \cite{Peyneau08}) amount $\Delta \phi$ for $\mu=\mu_1$. As $\mu$ is changed adiabatically, avalanches will be triggered when the force in a contact vanishes, as numerically investigated in \cite{Combe00}.
There must be of the order $N_a\sim \mu_1/\delta\mu_N\sim 1/\delta\mu_N$ avalanches between $\mu=0$ and $\mu_1$. Thus an avalanche leads to an average change of packing fraction  $\langle \delta \phi_N\rangle  \sim\Delta \phi/ N_a\sim \delta \mu_N$, i.e.  Eq.(\ref{21}). 
 
Note that Eqs.(\ref{10},\ref{13},\ref{21}) imply a relationship between coordination and packing fraction:
 \eq{ \label{coord}
  \delta z \sim \delta \phi^{1/y_\mu},
 }
 with $1/y_\mu \approx 0.83$. 



\section{Viscous to Inertial Transition}
\label{S4a}

Experiments \cite{Fall10}  and simulations \cite{Vagberg14,Trulsson12,Maiti14,Kawasaki14} report a transition from viscous to inertial flows as the strain rate $\dot \epsilon$ is increased at fixed packing fraction. For hard frictionless particles, the location of this transition can be computed precisely in our framework. The total power dissipated ${\cal P}_{tot}$ has a contribution from the viscous drag (${\cal P} \propto  N \eta_0 {\cal L}^2\dot\epsilon^2$, as discussed in the introduction) and from collisions (${\cal P} \propto N \dot\epsilon E_c/\epsilon_v\sim N \dot\epsilon^3 M {\cal L}^4$, where we used $\epsilon_v\sim 1/{\cal L}^2$). Thus one gets
\be
\label{p}
{\cal P}_{tot}\approx N C_1 \eta_0 D^{d-2}{\cal L}^2\dot\epsilon^2 + N C_2  M\dot\epsilon^3 {\cal L}^4,
\ee
leading to a crossover strain rate  $\dot\epsilon_{v\rightarrow i}$ above which dissipation is dominated by collisions:
\be
\label{co}
\dot\epsilon_{v\rightarrow i}\propto \frac{\eta_0 D^{d-2}}{M{\cal L}^2}\propto \frac{\eta_0 D^{d-2}}{M} \delta \phi^{1/\gamma_\phi}\sim  \frac{\eta_0}{ D^{2}\rho}\delta \phi^{2.83}
\ee
The stress scale $\sigma_{v\rightarrow i}$ at which this cross over occurs is thus  $\sigma_{v\rightarrow i}\sim \eta(\phi) \dot\epsilon_{v\rightarrow i}\sim\eta_0^2/( D^2 \rho)$, which is independent of $\phi$.  Thus the regime of strain rate where inertia is negligible vanishes rapidly when the jamming transition is approached \cite{Maiti14}.
Equating the total power dissipated of Eq.(\ref{p}) with the  power injected ${\cal P}\sim \Omega \sigma \dot\epsilon$, one gets the following scaling form for the viscosity $\eta\equiv \sigma/\dot\epsilon$:
\be
\label{vis}
\frac{\eta}{\eta_0}= \delta \phi^{-1/\gamma_\phi} f \left(\frac{\dot\epsilon}{\dot\epsilon_{v\rightarrow i}} \right)
\ee
where the scaling function $f$ satisfies $f(x)\sim x^0$ as $x\rightarrow0$ and $f(x)\sim x$ as $x\rightarrow\infty$. 


%


\begin{widetext}
\begin{table}[ht]
\begin{tabular}{| c | c | c || c  | c | c |}
\hline
\; Regime \; & \; Relation \; & \; Prediction \; & Experiment & Frictionless Sim'n & \; Frictional Sim'n \; \\
\hline \hline
 &  $\delta \mu \sim \Ic^{\alpha_\mu}$ & $\alpha_\mu = 0.35$  & 1 \cite{Forterre08} & 0.38(4) \cite{Peyneau08} & 0.81(3) \cite{Peyneau09}, 1 \cite{Bouzid13}, 1 \cite{Azema14}, 1 \cite{Trulsson12}\\   
Inertial & $\delta \phi \sim \Ic^{\alpha_\phi}$ & $\alpha_\phi = 0.35$   & 1 \cite{Forterre08} & 0.39(1) \cite{Peyneau08}  & 0.87(2) \cite{Peyneau09}, 1 \cite{Azema14}, 1 \cite{Trulsson12} \\  
& $\delta \mu \sim N^{-\alpha_N}$ & $\alpha_N = 1.19$ & & 1.16(4) \cite{Combe00} & \\  
 & $\Lc \sim \Ic^{-1/2}$ & $1/2$ & 0.7 \cite{Pouliquen04}, 0.7 \cite{MiDi04} & 0.48 \cite{Peyneau08} & 0.5 \cite{MiDi04} \\ 
\hline
  & $\eta \sim |\delta \phi|^{-1/\gamma_\phi}$ & $\gamma_\phi^{-1} = 2.83$ & 2 \cite{Boyer11}, 2 \cite{Ovarlez06}&  \parbox[c][1cm]{4cm}{2.6(1) \cite{Olsson12}, 2.77(20) \cite{Olsson11},  \\ 2.2 \cite{Andreotti12}, 2.5 \cite{Peyneau09}, \underline{2.77} } & \\  
Viscous & $\delta \mu \sim \Jc^{\gamma_\mu}$ & $\gamma_\mu = 0.35$ & 0.38 \cite{Lespiat11}, 0.42 \cite{Cassar05,Lespiat11}, 0.5 \cite{Boyer11} & 0.37 \cite{Peyneau09},  0.25 \cite{Olsson11}, \underline {0.32} & 0.5 \cite{Trulsson12} \\  
& $\delta z\sim {\cal J}^{\gamma_z}$ &$\gamma_z=0.30$ & &\underline{0.30}& \\
 & $\ell_c \sim |\delta \phi|^{-\gamma_\ell/\gamma_\phi}$ & $\gamma_\ell/\gamma_\phi = 0.43$ & & 0.6(1) \cite{Olsson07} & \\  
\hline
 & $d\Lc/d\gamma \sim -\Lc^{3}$ & $3$ &  & \underline{3} & \\
General & $\epsilon_v \sim \Lc^{-2} \sim \Jc $ & (-2,1) &  & $\epsilon_v \sim \underline{\Lc^{-2}}$, $\epsilon_v \sim \Jc$ \cite{Olsson10a} & \\
& $\edot_{v\rightarrow i} \sim \delta \phi^{\gamma_{\edot}}$ & $\gamma_{\edot} = 2.83$ & 1 \cite{Fall10} & & \\
\hline
\end{tabular}
\caption{Predicted critical exponents {\it vs.} values from experiments and numerical simulations, with and without frictional interactions. Underlined values correspond to the simulations presented in this paper. The values extracted in Ref.\cite{Peyneau09} correspond to simulations closest to hard spheres (the ``roughness parameter" of that reference is $10^{-4}$). When available, error bars are indicated by the notation $0.38(4) = 0.38 \pm 0.04$, $2.77(20)=2.77 \pm 0.20$, etc. }
\end{table}

\end{widetext}

\section{Comparison with observations}
\label{S6}
\subsection{Suspension Flows} 

{\it Simulations}: Scaling behavior described by Eqs.(\ref{03},\ref{04},\ref{06}) has been precisely characterized in simple numerical models of suspension flow, in particular for frictionless particles \cite{Olsson07,Olsson11,Olsson12,Andreotti12,Lerner12a}. The divergence of viscosity yields an exponent $1/\gamma_\phi \in [2.5,2.8]$ for the most recent data with the largest system size, in quantitative agreement with our prediction $1/\gamma_\phi = 2.83$. The exponent  $\gamma_\mu$ characterizing the stress anisotropy lies between $\gamma_\mu\in [0.25, 0.37]$ consistent with our prediction $\gamma_\mu=0.35$. Measurements of the correlation length exponent in terms of the packing fraction are scarce and not very recent (see \cite{During14} for measurement of length scale {\it vs} coordination), and it would be valuable to have more accurate measurements. Olsson and Teitel reported $\gamma_\ell/\gamma_\phi=0.6(1)$, in reasonable agreement with our prediction $\gamma_\ell/\gamma_\phi = 0.43$. Coordination was measured in a quasistatic simulation using soft particles \cite{Heussinger09}, finding $\delta z \sim \delta \phi$. This observation, which was performed over a very limited range, is consistent with our prediction $\delta z \sim \delta \phi^{0.83}$ (Eq.\ref{coord}).

\begin{figure*}[t!] 
\includegraphics[width=\textwidth,clip]{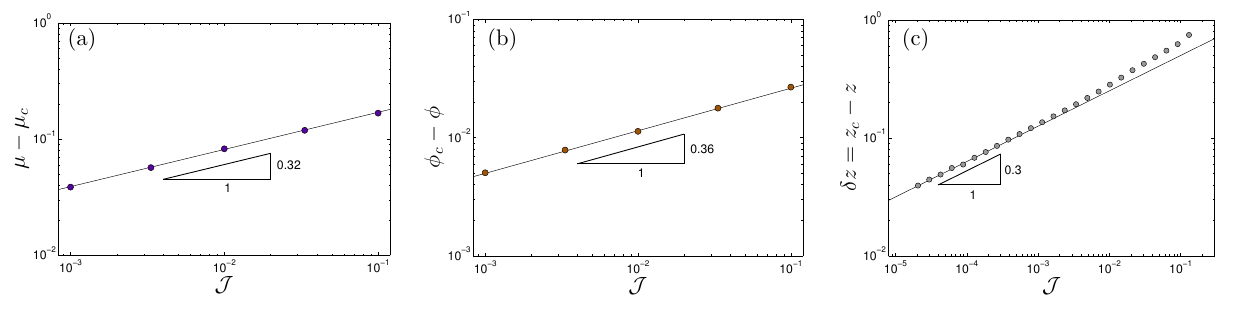}
\caption{(Color online) Numerical verification of scaling relations in the ASM. (a) Stress anisotropy $\mu$ vs. viscous number $\Jc$. Theory predicts an exponent $0.35$. (b) Volume fraction $\phi$ vs. viscous number $\Jc$. Theory predicts an exponent $0.35$. (c) Coordination deficit $\delta z$ vs. viscous number $\Jc$. Theory predicts an exponent $(1+\theta_e)/(4+2\theta_e) \approx 0.30$. 
}\label{fdmu}
\end{figure*}

\begin{figure*}[t!] 
\includegraphics[width=\textwidth,clip]{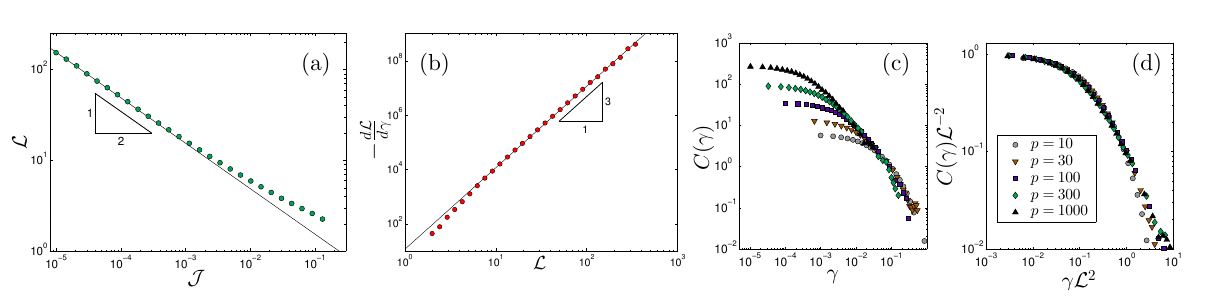}
\caption{(Color online) Numerical verification of kinematic scaling relations in the ASM. (a) Lever amplitude $\Lc$ vs. viscous number $\Jc$. Theory predicts an exponent $-1/2$. (b) Relaxation of lever amplitude in between collisions. Theory predicts $d\Lc/d\gamma \sim -\Lc^3$. (c,d) Autocorrelation of relative velocities $C(\gamma)$ vs. strain $\gamma$. Panel (d) shows a scaling collapse of $C(\gamma)$, indicating the presence of a strain scale $\epsilon_\eta \sim \Lc^{-2}$, as predicted. }\label{fell}
\end{figure*}

To supplement these results, and to show that we correctly describe the scaling behavior of microscopic observables not easily measured in experiments, we have performed simulations in a simple model of suspension flow. This model is used by various authors, and is a variant of the bubble model of Durian, except that particles are hard. We dubbed it the Affine Solvent Model (ASM) \cite{Lerner12a}, as in this model the solvent is assumed to flow in an affine way, unperturbed by the particles. Thus hydrodynamic interactions are neglected, and damping occurs when particles move with respect to the solvent. Observations indicate that the singular behavior is preserved when more realistic lubrication forces are considered \cite{Andreotti12, Vagberg14b,Peyneau08}, in agreement with our framework. We simulated steady-state shear of 50:50 binary mixtures of $N=1000$ particles in three dimensions, with the ratio of diameters of the large and small particles chosen to be 1.4. We collected data under both constant pressure and constant volume setups; see \cite{Lerner13} for details about the simulation methods. Our most accurate results on the exponents $\gamma_\mu$ and $\gamma_\phi$, shown in Fig.~\ref{fdmu}ab, give exponents $\gamma_\mu=0.32$ and $\gamma_\mu=0.36$, within error bars of our prediction $0.35$. In Fig.~\ref{fdmu}c we show that the coordination deficit $\delta z \sim \Jc^{0.30}$ is also quantitatively predicted. 

In Fig.~\ref{fell} we show the fundamental relations between microscopic quantities $\Lc, \Jc, \epsilon_v$ and $d\Lc/d\gamma$. As predicted, these show that in the ASM $\Lc \sim \Jc^{-1/2}$ and $d\Lc/d\gamma \sim \Lc^3$. In Fig.~\ref{fell}c we show the autocorrelation function $C(\gamma)= \langle V_r(0) V_r(\gamma) \rangle$, a function of shear strain $\gamma$, for various values of dimensionless pressure $p=1/\Jc$. These data are collapsed in Fig.~\ref{fell}d by plotting vs $\gamma \Lc^2$, indicating that relative velocities lose their memory after a strain scale $\epsilon_v \sim \Lc^{-2}$, as predicted. 
 
 In the literature, a wide variety of drag models have been considered. To be in the overdamped universality class, motion along floppy modes must be damped. When the drag is purely associated to longitudinal motion between particles in contact, or nearly in contact, the flow curves depend on the gap cutoff below which this drag is applied: inertial when only touching particles dissipate energy \cite{Vagberg14}, and viscous otherwise \cite{Maiti14}. This observation, which has been interpreted as a failure of universality, is natural from the present approach, since motion transverse to contacts dominates near jamming. 
 Thus model $h_c=0$ in \cite{Maiti14} and model CD$_n$ in \cite{Vagberg14} are inertial, while the other models considered therein are viscous.


{\it Experiments:} Experiments on frictionless hard particle systems near jamming are scarce, but this regime is accessible in foams. Foams are good systems to test result on hard spheres, as long as the shear stress is not sufficient to deform them significantly. In inverse avalanches it was that observed $\gamma_\mu = 0.38$ \cite{Lespiat11}, consistent with our prediction $\gamma_\mu = 0.35$.   

\begin{figure}[h!]
\includegraphics[width=0.40\columnwidth]{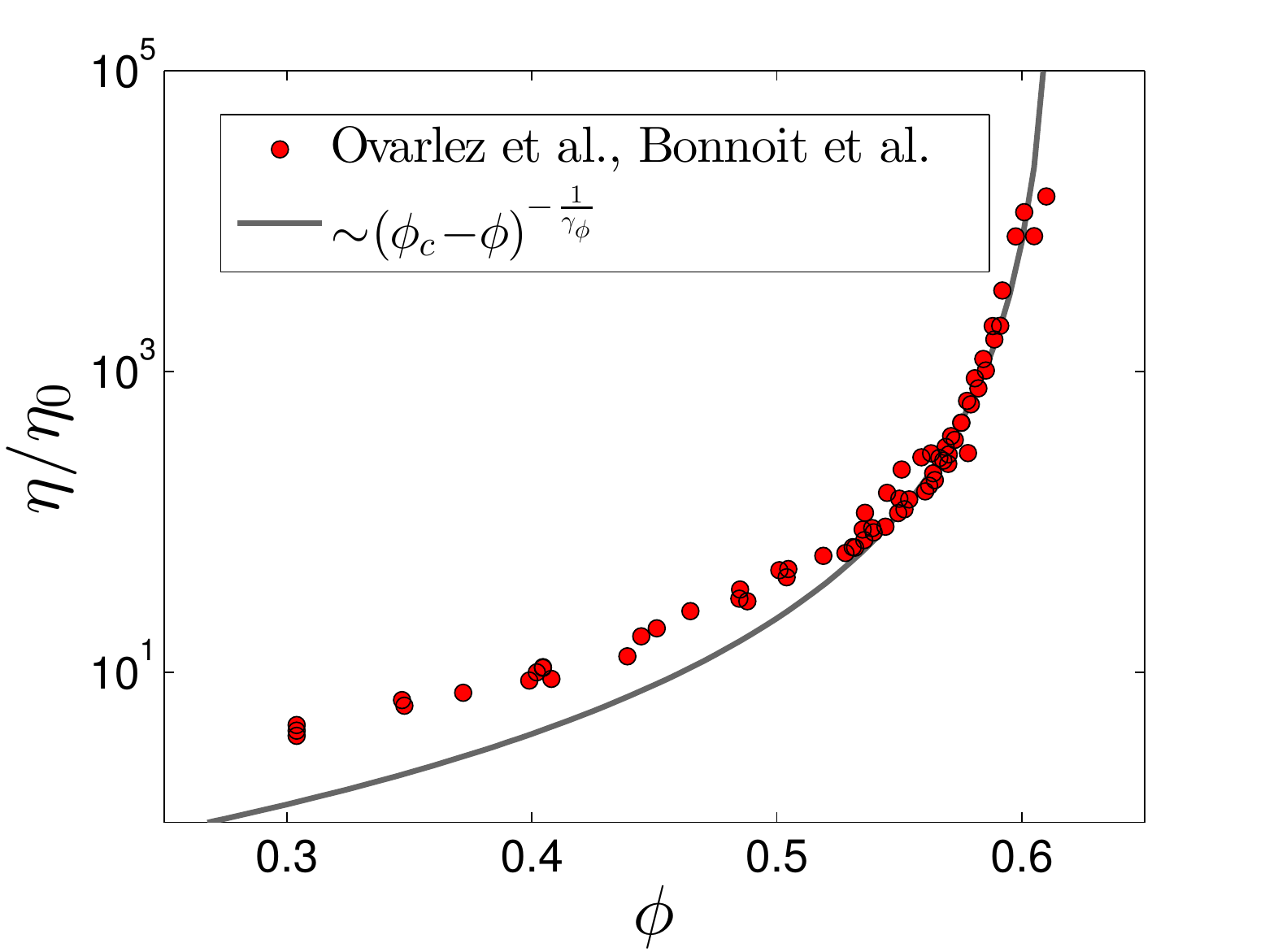}
\caption{Divergence of suspension viscosity as measured in experiments of Ovarlez et al \cite{Ovarlez06} and Bonnoit et al \cite{Bonnoit10} (symbols), compared to our prediction $\eta \sim \delta \phi^{-2.83}$ (solid). In these works the best-fit exponent was  $\approx 2$ when fitted on a large range of packing fraction. However, the prediction $2.83$ appears consistent with data close enough to $\phi_c$. }\label{f2}
\end{figure}

Most experiments are done with grains, i.e. frictional particles. It is often reported that the divergence of viscosity give an exponent $1/\gamma_\phi \approx 2$ \cite{Boyer11,Ovarlez06}, which may differ from our prediction $2.83$. However, a re-analysis of the data, shown in Fig.~\ref{f2}, suggests that the exponent 2 may simply reflect corrections to scaling: $2.83$ appears to work for $\phi \gtrsim 0.53$. Measurements of the exponents $\gamma_\mu$  yield $\gamma_\mu = 0.42$ \cite{Cassar05,Lespiat11},  rather close to our prediction $\gamma_\mu=0.35$. Frictional simulations suggest $\gamma_\mu = \gamma_\phi = 0.5$ \cite{Trulsson12}, but these measurements are lacking error bars. Thus for non-Brownian suspensions more accurate measurements are required to decide if frictional and frictionless particles behave identically, or not.


\subsection{Dry Flows} 

Dry granular flows, where inertia dominates, have been simulated with varying degrees of realism \cite{Combe00,Hatano07,Peyneau08,Peyneau09,Otsuki09,Heussinger09,Otsuki11,Chialvo12}. In a simple quasistatic ($\Ic=0$) model using hard frictionless particles poised near isostaticity \cite{Combe00}, Combe and Roux have measured explicitly the system-size dependence of stress increments needed to cause instability. The result is $\delta \mu \sim N^{-\alpha_N}$ with $\alpha_N = 1.16(4)$, in very good agreement with our prediction $\alpha_N = 1.19$.

At finite $\Ic$, but still with frictionless particles, Refs.(\cite{Peyneau08,Peyneau09}) measured the exponents $\alpha_\mu$ and $\alpha_\phi$, finding $\alpha_\mu=0.38(4)$ and $\alpha_\phi=0.39(1)$, again apparently in quantitative agreement with our prediction $\alpha_\mu=\alpha_\phi=0.35$. Moreover, assuming that relative velocities scale as non-affine velocities,  our prediction on $\Lc$ also appears correct: $\Lc \sim \Ic^{-0.48}$ is observed in \cite{Peyneau08}, in agreement with our prediction $\Lc \sim \Ic^{-1/2}$. Finally, although these authors have observed that the coordination converges to isostaticity at jamming, no exponent is reported. Measuring coordination precisely in the inertial regime may, however, be more difficult to perform than in the viscous case  \footnote[10]{When the restitution coefficient is not small, particles pushed against each other may exchange momentum by bouncing very rapidly on each other before forming a permanent contact. To define a contact precisely one may thus need to perform some kind of time-averaging, in a spirit similar to what is done in colloidal systems \cite{Brito06,Brito09}. }.

Friction appears to have a strong effect on dry inertial flow: simulations with Coulomb friction coefficients $\approx 0.5$ indicate that $\alpha_\mu \approx \alpha_\phi $ $\in (0.8, 1)$ \cite{Peyneau09,Bouzid13,Azema14,Trulsson12}, definitely distinct from our prediction $\alpha_\mu=\alpha_\phi=0.35$. In experiments on dry granular flow, it is likewise found that $\alpha_\mu \approx \alpha_\phi \approx 1$ \cite{Forterre08}. These data suggest that the present theory needs to be modified for frictional particles, at least in the inertial case. Consistent with these remarks, the viscous-to-inertial crossover is observed to satisfy $\gamma_{\edot} \approx 1$ in a frictional system \cite{Fall10}, off from our prediction $\gamma_{\edot} = 2.83$. 
Nevertheless our prediction on velocity fluctuations appears to be more accurate, as non-affine velocities are found to scale as 
$ \Ic^{-0.5}$  in simulations \cite{MiDi04} as we predict for relative velocities. In experiments the reported exponent is slightly larger, $\approx 0.7$ \cite{Pouliquen04,MiDi04}.





\section{Conclusion}

\subsection{Summary}
In a first step, we related the power dissipated in flow of frictionless particles to certain microscopic kinetic quantities. The latter control singularities in the rheological properties near jamming. In a second step, we have computed these quantities, using  a perturbation around the solid phase. Our main hypothesis is that
configurations in flow are similar to jammed configurations at maximum stress anisotropy $\mu_c$, destabilized by an 
additional stress increment $\delta \mu$. In this approach, the properties of the solid phase are central, in particular the
fact that  the density of contacts which can couple to external forces is singular at small forces, and characterized by a non-trivial exponent $\theta_e$. 
Our description of flow can thus be thought as that of a jammed solid, populated by elementary excitations corresponding to the opening of weak contacts, of density $\delta z$. 

Our work is part of a more general approach seeking to describe in real space the excitations that govern particulate materials and  their response. The  excitations studied here are associated to the rewiring of the contact network. Beyond flow, we have argued previously that these excitations control the stability of the solid phase: if the exponent $\theta_e$ were smaller, packings would collapse and have extensive rearrangements as soon as they are perturbed \cite{Wyart12,Lerner13a,Muller14}. Thus the value $\theta_e$ is fixed by stability constraints in the solid phase, and in turn affects flow properties.  
A similar situation occurs for soft vibrational modes, which are known to be present in amorphous solids (where they are referred to as the boson peak). We have argued \cite{Wyart05a, Brito06,Brito09,DeGiuli14b} that the structure of amorphous solids near jamming is such that soft vibrational modes are  stable, but barely so, a view also supported by recent calculations in infinite dimensions \cite{Charbonneau14a, Charbonneau14}. Once again this situation leads to a singular density of excitations (in that case the density of vibrational modes $D(\omega)$), causing anomalous elastic and transport properties \cite{DeGiuli14,DeGiuli14b}. Similar cases where stability is marginally satisfied and where the  density of excitations is singular occur in other glassy systems (such as spin and Coulomb glasses), and is expected if interactions are sufficiently long-range  \cite{Muller14}.



\subsection{Some open questions}
Although we believe that our assumption on the nature of flowing configurations is essentially correct, it would be very valuable to justify it from purely dynamical considerations. Work in that direction is in progress \cite{DuringNS}.  Another challenge concerns length scales. Physically the length $\ell_{n.l.}$ that characterizes non-local effects in flow when a boundary is present \cite{Bouzid13,Pouliquen09,Staron10,Henann13} is visible in many experiments, and of practical importance. It is presently unclear if this  length scale corresponds to $\ell_c$, which characterizes the main decay of the velocity correlation function. Indeed other length scales can be defined in flow \footnote[3]{We have argued previously that an additional length scale characterizes the far-tail decay of the velocity correlations, going as $\ell_r\sim 1/\sqrt{\cal J}$ \cite{During14}. A third length scale $\ell^*\sim \ell_c^2$ characterize how pinning the system at a distance rigidifies it \cite{Wyart05,During13,Schoenholz13}.   To which of the three length scales described above $\ell_{n.l.}$ is associated to, if any, is presently unclear.}. This question would benefit from more accurate measurements.

Finally, one central remaining  question is the  role played by friction.  As discussed in Section \ref{S6}, and visible in Table 1, friction  strongly affects critical exponents in the inertial case, but only weakly, if at all, in the viscous case. One key assumption of our approach, proximity to isostaticity, appears to be valid in the problematic inertial case: under constant stress boundary conditions, the steady state of simple shear dry granular flow is very nearly isostatic \cite{Kruyt10}. In our view, a central question for the future is what controls the stability of such isostatic frictional systems, how these respond to an additional stress anisotropy $\delta \mu$, and how the combination of finite softness, inertia and friction qualitatively affects the flow curves \cite{Otsuki11}.

 
 \acknowledgements
 
We thank B. Andreotti, M. Cates, Y. Forterre, J. Lin, B. Metzger, M. Mueller, O. Pouliquen, A. Rosso, L. Yan, and F. Zamponi for discussions and B. Andreotti for providing the compilation of data used in  Fig.(\ref{f2}). 
MW acknowledges support from NSF CBET Grant 1236378, NSF DMR Grant 1105387, and MRSEC Program of the NSF DMR-0820341 for partial funding. GD~acknowledges support from CONICYT  PAI/Apoyo al Retorno 82130057.

\bibliography{../../bib/Wyartbibnew,../bib/Wyartbibnew}

\section{Appendix}

\subsection{Virtual Work Theorems}

\newcommand{\ten}[1]{\overset{\text{\tiny$\bm\leftrightarrow$}}{#1}}

In the main text we make use of two work theorems, which we derive here in the frictionless case \cite{Roux00}. We also derive the microscopic expression for the stress tensor. 

The first work theorem applies to any jammed packing ($z>z_c$) and begins with the equations for force balance,
\eq{ \label{fb}
\vec{F}^{ext}_i = \sum_{ij} f_{ij} \vec{n}_{ij},
}
where $\vec{F}^{ext}_i$ is the external force on particle $i$, and $f_{ij}$ is the contact force in contact $ij$. Contracting this equation along an arbitrary (``virtual") displacement field $\delta \vec{R}_i$, and summing over all particles, we find
\eq{
\sum_i \delta \vec{R}_i \cdot \vec{F}^{ext}_i & = \sum_i \delta \vec{R}_i \cdot \sum_{ij} f_{ij} \vec{n}_{ij} \notag \\
& =- \sum_{ij} f_{ij} \delta r_{ij},
}
where $\delta r_{ij}=(\delta \vec{R}_j-\delta \vec{R}_i) \cdot \vec{n}_{ij}$ is the normal displacement at contact $ij$. This statement is called the theorem of virtual work. When a packing is jammed, there exists solutions to \eqref{fb} where all the forces are applied at the boundary of the packing. In this case we identify $\sum_i \delta \vec{R}_i \cdot \vec{F}^{ext}_i=W$ as the work injected in the displacement $\{ \delta \vec{R}_i \}$. The work can be written $W = \Omega \ten{\sigma} : \ten{\epsilon} = -\Omega d \; p \epsilon_V + \Omega \sigma \epsilon$, where $\ten{\sigma}$ is the stress tensor, $\ten{\epsilon}$ is the strain tensor, $\epsilon$ is shear strain, and $\epsilon_V$ is the volumetric strain (zero if constant-volume boundary conditions are imposed, and positive for dilation). Therefore for a jammed packing
\eq{
W = - \sum_{ij} f_{ij} \delta r_{ij}.
}

A similar relation holds in an unjammed system. When $z<z_c$, there are floppy modes that pervade the system, i.e., any velocities imposed at boundaries can be accommodated by motions that maintain all contacts. Let us suppose that the list of contacts $\{ ij \}$ includes some contacts with walls at the boundary of the domain. The previous statements imply that the set of equations
\eq{ \label{fm}
u_{ij} = ( \vec{V}_j- \vec{V}_i) \cdot \vec{n}_{ij}
}
has a solution for the velocities $\{ \vec{V}_i \}$ when the $u_{ij}$ are nonzero only at boundaries. Contracting these equations of ``geometric balance" along an arbitrary force field gives
\eq{ \label{fm2}
\sum_{ij} f_{ij} u_{ij} = -\sum_i \vec{V}_i \cdot \sum_{ij} \vec{n}_{ij} f_{ij} = \sum_i  \vec{V}_i \cdot \vec{F}_i,
}
where $\vec{F}_i$ is the vectorial sum of contact forces incident on $i$; this is the theorem of complementary virtual work. In this case the LHS of \eqref{fm2} is nonzero only at boundaries, and inspection of this at a contact $ij$ shows that if the $\{ f_{ij} \}$ are taken as the true contact forces in contacts, then this is the power injected in the imposed velocity field $\{ u_{ij} \}$. Therefore the LHS is the power ${\cal P}$. 

Finally, let us show how the microscopic expression for the stress tensor can be obtained. In the jammed case, we multiply \eqref{fb} with the particle positions $\vec{R}_i$ and sum the resulting the equations (without contracting the vectors). This gives
\eq{ \label{stress1}
\sum_i \vec{F}^{ext}_i \vec{R_i} = - \sum_{ij} f_{ij} \vec{n}_{ij}  \vec{r}_{ij},
}
where $\vec{r}_{ij}= \vec{R}_j - \vec{R}_i$. The LHS of \eqref{stress1} is a discretization of a boundary integral $-\int_{\partial \Omega} \vec{n} \cdot \ten{\sigma} \vec{r} \; dS$, where $\vec{n}$ is an outward-facing normal to the boundary. By the divergence theorem, this is equal to $-\int_\Omega \nabla \cdot (\ten{\sigma} \vec{r} \; ) \; dV$. But then force balance implies $\nabla \cdot \ten{\sigma}=0$ so that $\int_\Omega \nabla \cdot (\ten{\sigma} \vec{r} \; ) \; dV = \int_\Omega \ten{\sigma}{}^t dV \equiv \Omega \ten{\sigma}{}^t$ and
\be
\label{A000}
\ten{\sigma} = \frac{1}{\Omega} \sum_{ij} f_{ij} \vec{r}_{ij} \vec{n}_{ij}.
\ee
Similar equations hold in the more general frictional case, including the effect of rotations \cite{Kruyt03}.

 \subsection{Shear modulus and anisotropy}
 In weakly coordinated packings of  elastic particles, generic elastic moduli are small and scale as  $\sim z-z_c$, a scaling that holds up to $\delta z=1/N$.
This is true except in the direction of the applied stress, where the modulus is large: it is not vanishing and goes as $(z-z_c)^0$ \cite{Wyart10b,Ellenbroek09}.
This result explains why the bulk modulus is always large for purely repulsive particles, as observed numerically \cite{Ohern03}. In an anisotropic packing carrying a shear stress $\sigma$ of order of the pressure $p$, the stiff mode of deformation is not a pure compression, as it now has a shear component. However, imposing some additional stress on the  system $\delta \sigma$ and $\delta p$ will generically couple to the soft  moduli, except if $\delta \sigma/\delta p=\sigma/p=\mu$. In our case we impose an additional shear stress increment with no additional compression:
there is thus a finite coupling to the weak elastic moduli leading to a large particle displacement, as we have assumed in the text to derive Eq.(\ref{8}).

\subsection{Derivation of $\delta\mu_N$ for strictly hard particles}
For a simple shear in the $xy$-plane, it is useful to write Eq.(\ref{A000}) in a compact notation as
\eq{
\sigma & = \frac{1}{\Omega} \langle f | l\rangle  \label{A001}\\
p & = \frac{d}{\Omega} \langle f | r\rangle,  \label{A002}
}
where $\sigma= \ten{\sigma}_{xy}$,  $|f\rangle$ is the vector of contact forces $f_{ij}$ (of dimension $N_c$, the number of contacts), $|r\rangle$ is the vector of the distances $r_{ij}$ between particles in contact, and $|l\rangle$ has components    $l_{ij}=({\vec r}_{ij}\cdot {\hat x}) ({\vec n}_{ij}\cdot {\hat y})$.
We denote by $|\delta f\rangle $ the change of contact forces induced by increasing the shear stress by $\delta \sigma$. It must obey the conditions:
\eq{
\frac{1}{\Omega}  \langle\delta f | l\rangle &=\delta \sigma\label{A003}\\
  \langle \delta f | r\rangle &=0\label{A004}
}
  
If  hard particles are compressed homogeneously from a loose state, say by reducing the linear size $L$ of a cubic box containing them,
the system will eventually jam into an isostatic configuration: there are just enough contacts to forbid floppy modes,
which involve the $Nd$ degrees of freedom of  the particles, as well as the dimension $L$ of the box. At that point, there is a single set of contact forces
that satisfies force balance on each particle, i.e. Eq.(\ref{fb}) with no LHS. However if the system is then allowed to shear (for example by deforming the square box into a rectangle), there is then one floppy mode associated to this additional degree of freedom, see e.g. \cite{Goodrich14}. It will disappear once a new contact is formed. At that point, the space of  contact forces satisfying force balance is of dimension two. This situation is  generic  in practical situations, for example when the shear stress is adiabatically increased to study plasticity in packings \cite{Combe00}.

 We denote by $|f_1\rangle$ and $|f_2\rangle$ an orthonormal basis of this space. The components of these vectors thus scale as $1/\sqrt{N}$. 
We choose $|f_1\rangle$ to be in the direction of the true contact forces  before the stress increment. Eq.(\ref{A002}) then implies for a purely repulsive system (where all contact forces must have the same sign) the following system-size dependence:
\ba
\langle f_1| r\rangle \sim \sqrt{N} \label{A005}\\
\langle f_1| l\rangle \sim \sqrt{N} \label{A006}
\ea
where the second relation stems from Eq.(\ref{A001}) and the assumption that $\sigma\sim p$, i.e $\mu\neq 0$. Our central assumption is that $|f_2\rangle$ is essentially a random vector with limited spatial correlations. More precisely we assume that:
\ba
\langle f_2| r\rangle \sim 1 \label{A007}\\
\langle f_2| l\rangle \sim 1 \label{A008}
\ea
as follows from the central limit theorem if the sums in Eqs.(\ref{A007},\ref{A008}) concerns weakly-correlated variables. 

Writing $|\delta f\rangle= \beta_1 |f_1\rangle +\beta_2 |f_2\rangle$,  one readily gets expressions for $\beta_1$ and $\beta_2$ from Eqs.(\ref{A003},\ref{A004}).
Using Eqs.(\ref{A005},\ref{A006},\ref{A007},\ref{A008}) one finds $\beta_1 \ll \beta_2$ and $\beta_2\sim \delta \sigma N$. We seek to compute the characteristic change of force in a contact $\delta f$, which then must follow:
\be
\label{A009} 
\delta f^2=\frac{\langle \delta f|\delta f \rangle}{N_c}=\frac{\beta_1^2+\beta_2^2}{N_c}\sim N\delta \sigma^2
\ee
which is equivalent to Eq.(\ref{8}).

\end{document}